\documentclass[aps,prl,reprint,groupedaddress]{revtex4-1}

\newcommand{\be}{\begin{equation}}
\newcommand{\ee}{\end{equation}}

\usepackage{verbatim}
\usepackage{latexsym,amssymb,float}
\usepackage{setspace}
\usepackage{graphicx}
\usepackage{epsfig}
\usepackage{amsmath}
\usepackage{bm}

\begin{document}

\title{Josephson Scanning Tunneling Spectroscopy in $d_{x^2-y^2}$-wave superconductors: a probe for the nature of the pseudo-gap in the cuprate superconductors}
\author{Martin Graham}
\author{Dirk K. Morr}
\affiliation{University of Illinois at Chicago, Chicago, IL 60607, USA}

\date{\today}

\begin{abstract}
Recent advances in the development of Josephson scanning tunneling spectroscopy (JSTS) have opened a new path for the exploration of unconventional superconductors. We demonstrate that the critical current, $I_c$, measured via JSTS, images the spatial form of the superconducting order parameter in $d_{x^2-y^2}$-wave superconductors around defects and in the Fulde-Ferrell-Larkin-Ovchinnikov state. Moreover, we show that $I_c$ probes the existence of phase-incoherent superconducting correlations in the pseudo-gap region of the cuprate superconductors, thus providing unprecedented insight into its elusive nature. These results provide the missing theoretical link between the experimentally measured $I_c$, and the spatial structure of the superconducting order parameter.
\end{abstract}


\maketitle

Visualizing the spatial structure of the order parameter in unconventional $d_{x^2-y^2}$-wave superconductors represents a crucial step towards discovering the microscopic origin of their complex properties. To achieve this goal, recent experimental efforts have focused on the development of Josephson scanning tunneling spectroscopy (JSTS) \cite{Sma01,Naa01,Rod04,Kim09,Sud14,Ham15,Jae16,Ran16}. The assumption underlying JSTS is that the Josephson current, $I_J$, \cite{Jos62} flowing between a superconducting JSTS tip and a superconductor is proportional to the local order parameter of the latter \cite{Amb63}, even if it varies on the length scale of a few lattice constants. A proof of this assumption, which until now has been lacking for unconventional superconductors \cite{Gra17}, would open unprecedented possibilities for the application of JSTS: it would allow one to gain insight not only into the response of the superconducting order parameter (SCOP) to defects and disorder, but also into its much anticipated form in the magnetic field induced Fulde-Ferrell-Larkin-Ovchinnikov (FFLO) state \cite{Ful64,Lar65,Bia03,Mat07,Radovan03}. Moreover, JSTS could be employed as a probe for superconducting correlations in the pseudo-gap (PG) region of the cuprate superconductors \cite{Sma01,Ber08,Jac16}, thus shedding light on the question of whether the PG arises from a pair-density-wave (PDW) \cite{Chen04,Berg09,Lee14,Fra15,Wang15} or phase-incoherent superconducting fluctuations \cite{Eme95,Fra98,Chen05,Wul09}. Indeed, recent JSTS experiment in the cuprate superconductor Bi$_2$Sr$_2$CaCu$_2$O$_{8+x}$ by Hamidian {\it et al.} \cite{Ham15} argued that the observed spatial oscillations in the critical current, $I_c$, provide evidence for the existence of a PDW. However, a complication in interpreting these experiments, and in identifying the relation between $I_c$ and the SCOP arises from the fact that the non-local nature of the SCOP in $d_{x^2-y^2}$-wave superconductors requires the use of spatially extended JSTS tips.

\begin{figure}[h]
\includegraphics[width=8cm]{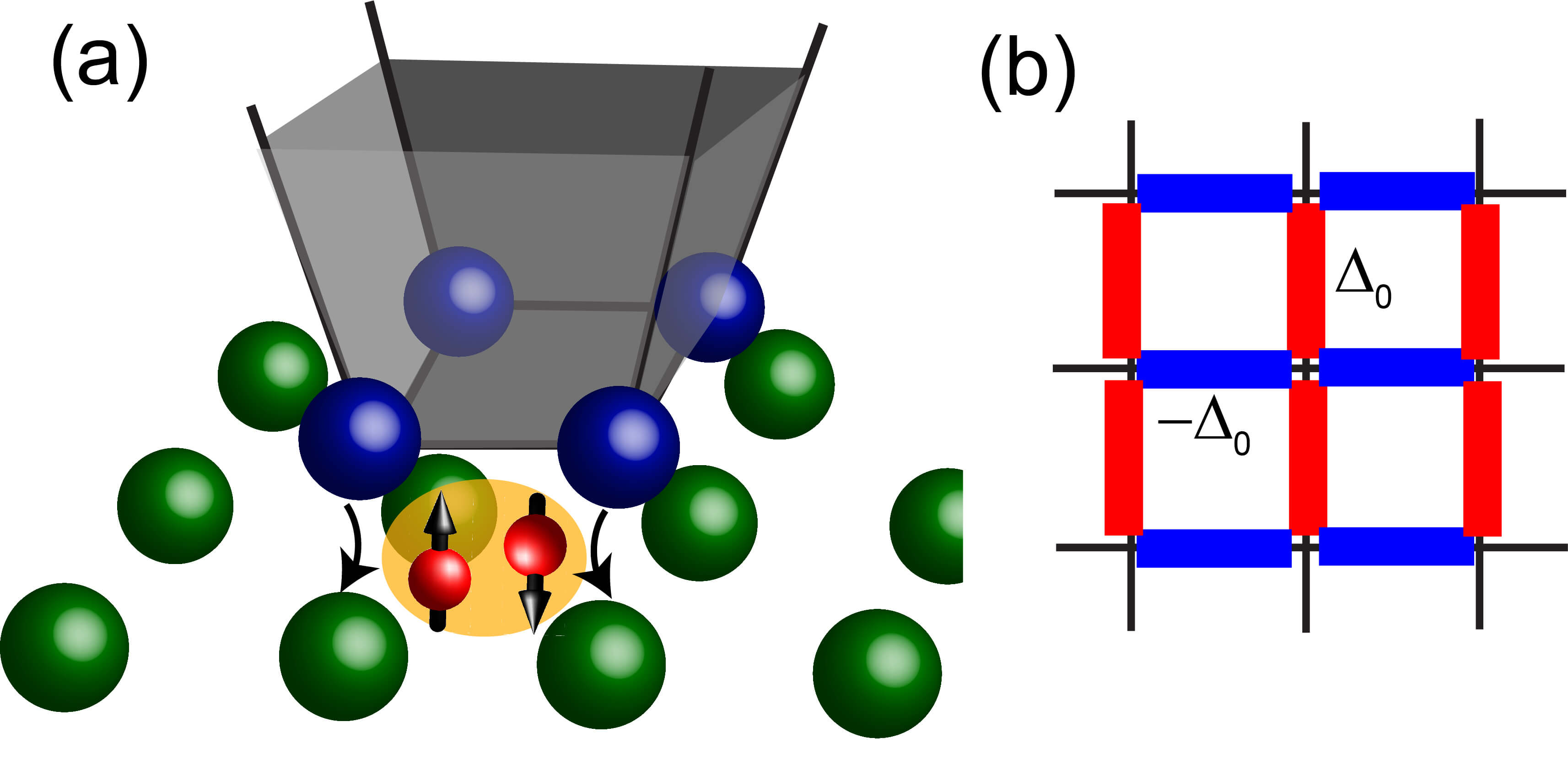}%
 \caption{(a) Schematic picture of Cooper-pair tunneling in a $d_{x^2-y^2}$-wave superconductor from a spatially extended $(2 \times 2)$ JSTS tip. (b) Spatial structure of the SCOP in a $d_{x^2-y^2}$-wave superconductor with red/blue bonds representing the SCOP $\pm \Delta_0$ between nearest-neighbor sites.}
 \label{fig:tip}
 \end{figure}

In this article, we provide the theoretical proof that the Josephson current measured via JSTS can provide direct insight into the spatial structure of the SCOP in $d_{x^2-y^2}$-wave superconductors, and into the existence of superconducting correlations in the PG regime of the cuprate superconductors. Using a Keldysh non-equilibrium Green's function formalism \cite{Kel65,Ram86}, we demonstrate that the Josephson current, $I_c$, flowing from a spatially extended superconducting JSTS tip with $d_{x^2-y^2}$-wave symmetry, into a $d_{x^2-y^2}$-wave superconductor [schematically shown in Fig.~\ref{fig:tip}(a)] images the spatial structure of the SCOP (averaged over an area of the size of the tip) in the latter. Thus, for sufficiently small tip sizes, it is possible to gain insight into the spatial structure of the SCOP on the atomic length scale.
This allows one to visualize the response of the SCOP to defects, and to image its long sought spatial structure, including a sign change, in the FFLO phase. Moreover, as a non-zero $I_c$ only requires the existence of local superconducting correlations and not of global phase coherence, we predict that a finite $I_c$ should be measured in the PG region of the cuprate superconductors if the latter arises from phase-incoherent superconducting pairing. These results demonstrate that JSTS possesses unprecedented potential for gaining insight into the spatial nature of strongly correlated, unconventional superconductivity.

The starting point for investigating the relation between the spatial form of the critical Josephson current and the SCOP in $d_{x^2-y^2}$-wave superconductors is the Hamiltonian $H=H_s + H_{tip} + H_{tun}$ where
\begin{align}
\label{eq:Hs}
H_s= &  -\sum_{{\bf r},{\bf r'}, \sigma} t_{{\bf r}{\bf r}^\prime} c_{{\bf r}\sigma}^\dagger c_{{\bf r}^\prime \sigma} -\mu\sum_{{\bf r} ,\sigma}c_{{\bf r} \sigma}^\dagger c_{{\bf r} \sigma} \nonumber \\
 &- \sum\limits_{{\bf r,r^\prime}} \left [ \Delta_{{\bf r}{\bf r}^\prime} c_{{\bf r}, \uparrow}^\dagger c_{{\bf r}^\prime, \downarrow}^\dagger + H.c.\right ] + U_0 \sum_{\sigma}c_{{\bf R} \sigma}^\dagger c_{ {\bf R}\sigma}  \nonumber \\
 & - \frac{g \mu_B}{2} {\bf H } \cdot \sum_{{\bf r}, \alpha,\beta} c^\dagger_{ {\bf r},\alpha} {\bm \sigma}_{\alpha,\beta} c_{ {\bf r},\beta}
\end{align}
Here, $-t_{{\bf r}{\bf r}^\prime}$ is the electronic hopping between sites ${\bf r}$ and ${\bf r}^\prime$ in the superconductor, $\mu$ is the chemical potential, and $c^\dagger_{{\bf r} \sigma}$ ($c_{{\bf r} \sigma}$) creates (annihilates) an electron with spin $\sigma$ at site ${\bf r}$. $\Delta_{{\bf r}{\bf r}^\prime}$ is the non-local (bond) SCOP with $d_{x^2-y^2}$-wave symmetry, which is non-zero only between nearest-neighbor sites and changes sign between the $x$- and $y$-directions, i.e, $\Delta_{{\bf r}{\bf r}^\prime} = \pm \Delta_0$, as shown in Fig.~\ref{fig:tip}(b) for a translationally invariant system. $U_0$ is the scattering potential of a non-magnetic defect located at ${\bf R}$ and the last term represents the Zeeman coupling in a Pauli-limited superconductor, necessary to create the FFLO phase \cite{Yan09,Liu12}. We employ a set of parameters that is characteristic of the cuprates' electronic structure with next-nearest-neighbor hopping $t^\prime/t = -0.4$, and $\mu/t=-1$.

To account for spatial oscillations of the SCOP, we compute it self-consistently via
\begin{align}
\Delta_{{\bf r}{\bf r}^\prime}=-\frac{V_{{\bf r}{\bf r}^\prime}}{\pi}\int_{-\infty}^\infty d\omega \, n_F(\omega) \text{Im}[F_{sc}({\bf r'}, \downarrow; {\bf r},\uparrow, \omega))] \label{eq:OP}
\end{align}
where $V_{{\bf r}{\bf r}^\prime}$ is the superconducting pairing potential between nearest-neighbor sites, $n_F(\omega)$ is the Fermi distribution function, and $F_{sc}$ is the non-local, retarded anomalous Green's function of the $d_{x^2-y^2}$-wave superconductor [see Supplemental Material (SM) Sec.~I]. We model the JSTS tip as a spatially extended $d_{x^2-y^2}$-wave superconductor with $(n_x \times n_y)$ sites [see Fig.~\ref{fig:tip}(a)], described by the Hamiltonian $H_{\rm tip}=H_{\rm tip}^{\rm n}+H_{\rm tip}^{\rm sc}$, where $H_{\rm tip}^{\rm n}$ represents the normal state electronic structure of the tip (see SM Sec.I), and
\begin{align}
H_{\textup{tip}}^{\textup{sc}}=-\sum_{{\bf r}, {\bf r}^\prime } \Delta^{\textup{t}}_{{\bf r} {\bf r}^\prime} d_{{\bf r} \uparrow}^\dagger d_{{\bf r}^\prime \downarrow}^\dagger + H.c.
\end{align}
where $\Delta^{\textup{t}}_{{\bf r} {\bf r}^\prime} = \pm \Delta_{tip}$ is the tip's superconducting $d_{x^2-y^2}$-wave order parameter, $d^\dagger_\sigma$ ($d_\sigma$) create (annihilate) an electron with spin $\sigma$ in the tip, and the sum runs over all tip sites. Finally, the tunneling Hamiltonian is given by
\begin{align}
H_{\rm tun}=-t_0\sum_{{\bf r}, \sigma} c^\dagger_{{\bf r} \sigma}d_{{\bf r} \sigma}+ H.c.
\end{align}
where ${\bf r}$ denotes sites both in the tip and the $d_{x^2-y^2}$-wave superconductor between which electrons can tunnel.

A DC Josephson current, $I_J$ \cite{Jos62}, between the JSTS tip and the superconductor arises from a phase difference, $\Delta \Phi $, between their SCOPs, which can be gauged away \cite{Cue96}, yielding real SCOPs and a phase-dependent tunneling amplitude $t_T =  t_0 e^{i \Delta \Phi/2}$. Using the Keldysh Green's function formalism \cite{Kel65,Ram86}, one obtains $I_J=I_J^\uparrow + I_J^\downarrow $ to lowest order in the hopping $t_T$ \cite{Cue96} as
\begin{align}
I^\sigma_{J} &= 4\frac{e}{\hbar} t_0^2 \sin{\left(\Delta \Phi \right)} \int \frac{d\omega}{2\pi}n_F(\omega) \nonumber \\
& \times \sum_{ {\bf r}, {\bf r}^\prime} \text{Im}[F_t({\bf r},\sigma;{\bf r'},{\bar \sigma}, \omega) F_{sc}({\bf r'}, {\bar \sigma}; {\bf r},\sigma, \omega)]
\label{eq:Ieq}
\end{align}
where $F_t$ is the retarded anomalous Green's function of the tip (see SM Sec.~I), and the sum runs over all sites ${\bf r, r'}$ in the tip and superconductor
that are connected by a tunneling element. Finally,  $I_J = I_c \sin{\left(\Delta \Phi\right)}$ with $I_c$ being the critical Josephson current. Note that while $\Delta_{{\bf r}{\bf r}^\prime}$ is non-zero for nearest-neighbor sites only, $F_{sc}$ is non-zero for further neighbor sites, which therefore need to be included in the summation in Eq.(\ref{eq:Ieq}).

\begin{figure}[h!]
\includegraphics[width=8cm]{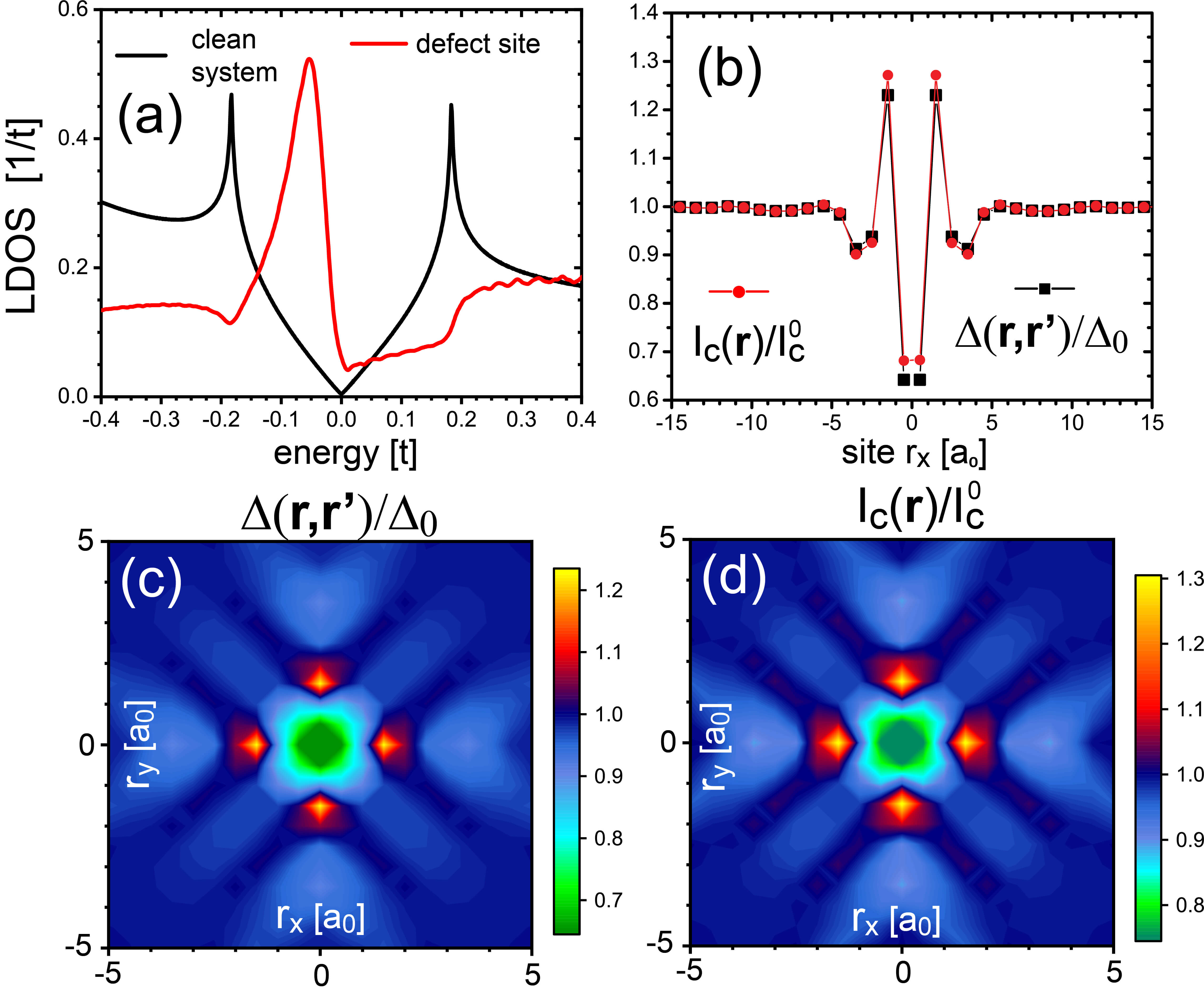}%
 \caption{(a) LDOS in a clean system and at the site of a defect with $U_0=1.5$t. (b) Normalized $\Delta({\bf r,r'})$ and $I_c$ for a $(2 \times 1)/(1 \times 2)$ tip plotted at $({\bf r}+{\bf r'})/2$ along a linecut through the defect site for $\Delta_{tip}=4\Delta_0$. $I_c^0$ is the Josephson current in the unperturbed system. (c) Spatial plot of $\Delta({\bf r,r'})$ and (d) $I_c$ for $\Delta_{tip}=\Delta_0$ and $T=0$.}
 \label{fig:OPJC}
 \end{figure}
Before discussing the characteristic features of $I_c$ in the pseudo-gap region of the cuprate superconductors, we first consider its hallmark signatures in a fully phase coherent $d_{x^2-y^2}$-wave superconductor. We begin by investigating the spatial form of the SCOP near a non-magnetic defect, which gives rise to the emergence of an impurity resonance in the local density of states (LDOS) \cite{Bal06,Hud99,Pan00,Bal95,Sal96,Sal97}, as shown in Fig.~\ref{fig:OPJC}(a). At the same time, the defect also induces spatial oscillations in the SCOP [see Figs.~\ref{fig:OPJC}(b) and (c)], which cannot be measured via conventional scanning tunneling spectroscopy \cite{Ham15,Ran16}. In Fig.~\ref{fig:OPJC}(b), we present the spatial form of the critical current, $I_c$, for a tip that consists of 2 sites, which is the smallest possible tip size  that still exhibits non-local $d_{x^2-y^2}$-wave correlations. The tip is aligned either along the $x$- or $y$-direction, representing a $(2 \times 1)$ or $(1 \times 2)$ tip, respectively.  The resulting $I_c$ probes the superconducting correlations between nearest-neighbor sites only, thus providing direct insight into the non-local bond SCOP.
The spatial form of $I_c$ for $\Delta_{tip}=4\Delta_0$ agrees very well with that of the SCOP [Fig.~\ref{fig:OPJC}(b)] implying that the spatial structure of an unconventional $d_{x^2-y^2}$-wave order parameter can be spatially imaged by the critical current. This good agreement is independent of the particular magnitude of the SCOP in the tip, as follows from a comparison of the spatial structure of the SCOP [Fig.~\ref{fig:OPJC}(c)] and of $I_c$ [Fig.~\ref{fig:OPJC}(d)] for $\Delta_{tip}=\Delta_0$. The wavelength of the oscillations along the $x/y$-axis both in $\Delta({\bf r, r^\prime})$ and $I_c$ is approximately 4$a_0$ [Fig.~\ref{fig:OPJC}(b)], which is close to that observed by Hamidian {\it et al.} \cite{Ham15}. This wavelength arises from scattering of electrons between the nearly parallel parts of the Fermi surface near $(0,\pm \pi)$ and  $(\pm \pi,0)$.

\begin{figure}[t]
\includegraphics[width=8cm]{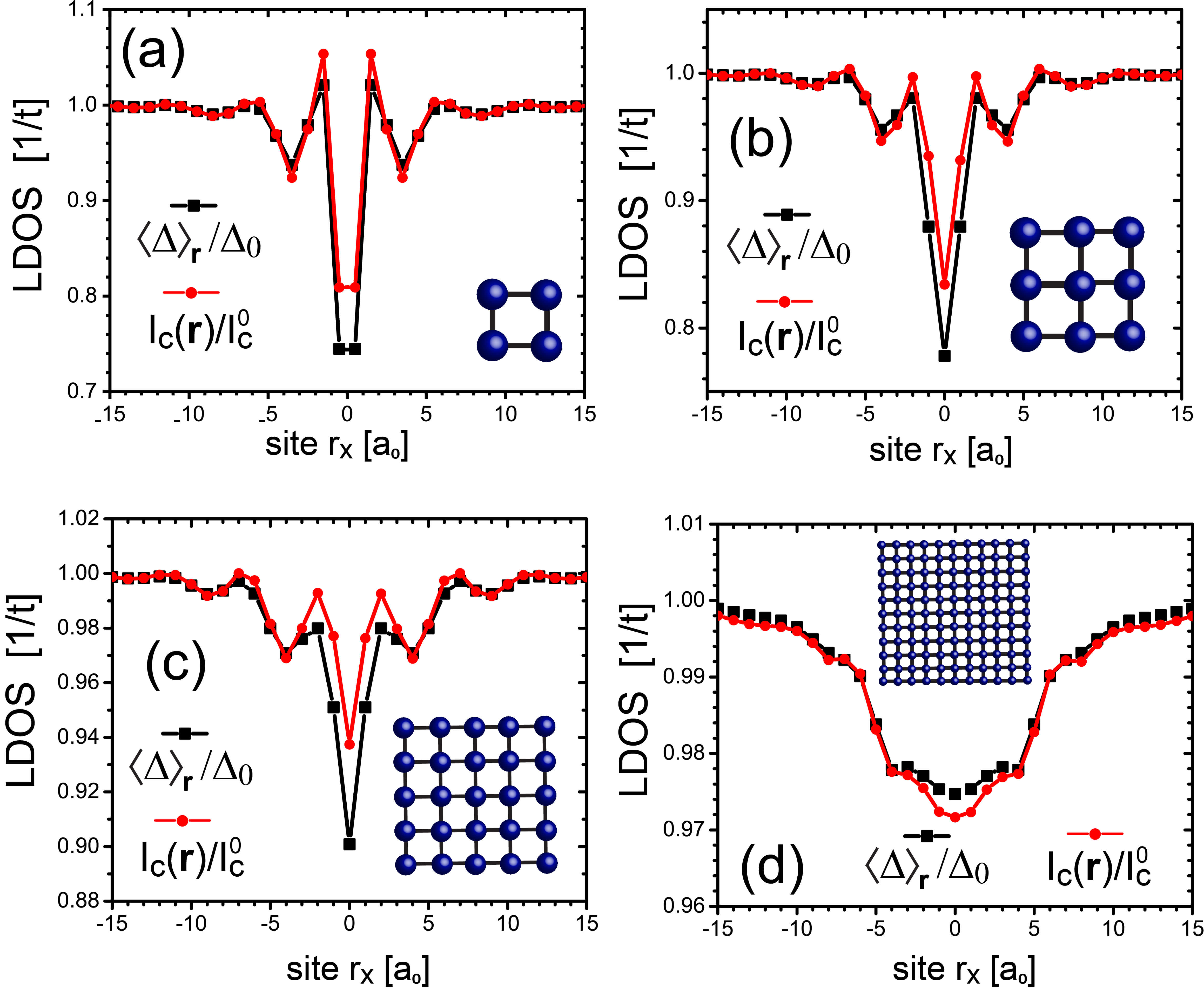}%
 \caption{Comparison of $I_c({\bf r})$, with {\bf r} being the center of the tip, and the spatially averaged SCOP $\langle \Delta \rangle_{\bf r}$ for (a) $(2 \times 2)$, (b) $(3 \times 3)$, (c) $(5 \times 5)$, and (d) $(11 \times 11)$ JSTS tips (the tip sizes are shown as insets). $\langle \Delta \rangle_{\bf r}$ is calculated by averaging the magnitude of $\Delta_{{\bf r}{\bf r}^\prime}$ over the region covered by the JSTS tip.}
 \label{fig:tip_size}
 \end{figure}

While the above results were obtained with the smallest possible tip size still exhibiting $d_{x^2-y^2}$-wave correlations, the JSTS tip employed by Hamidian {\it et al.} \cite{Ham15}, was created by picking up a nanometer-sized flake of Bi$_2$Sr$_2$CaCu$_2$O$_{8+x}$ with a tungsten tip. This immediately brings into question to what extent the critical current $I_c$ measured by such a spatially extended JSTS tip can still image the ``local" SCOP. To investigate this crucial question, we compare in Fig.~\ref{fig:tip_size} the critical current measured by several spatially extended JSTS tips of different sizes, with the SCOP averaged over the area covered by the tip, $\langle \Delta \rangle_{\bf r}$.
We find as expected that with increasing tip size, the agreement between the $I_c$ and the bond SCOP between nearest neighbor sites (at the center of the tip) worsens [cf., for example, $\Delta({\bf r,r'})$ in Fig.~\ref{fig:OPJC}(b) with $I_c$ in Fig.~\ref{fig:tip_size}(c)]. However, the agreement between the spatial form of $I_c$ and the averaged SCOP  $\langle \Delta \rangle_{\bf r}$ remains very good, as shown in Fig.~\ref{fig:tip_size}. Moreover, even for a large $(5 \times 5)$ tip [Fig.~\ref{fig:tip_size}(c)] (which is approximately the size of flake of Bi$_2$Sr$_2$CaCu$_2$O$_{8+x}$ used by Hamidian {\it et al.} \cite{Ham15}) the $\lambda=4a_0$ spatial oscillations are still visible. As with increasing tip size, the contribution to $I_c$ from nearly unperturbed areas increases even when the tip is centered above the defect, the relative spatial variation of $I_c$ around the defect becomes weaker, as follows from Figs.~\ref{fig:tip_size}(a)-(d). Thus, while with increasing tip size, $I_c$ does not any longer image the spatial structure of the bond order parameter, $\Delta({\bf r, r^\prime})$, it nevertheless provides insight into the form of the spatially averaged SCOP. We note that this result is largely robust against disorder in the tunneling amplitude (see SM Sec.~II), and thus also holds for disordered tips.

\begin{figure}[h]
\includegraphics[width=8cm]{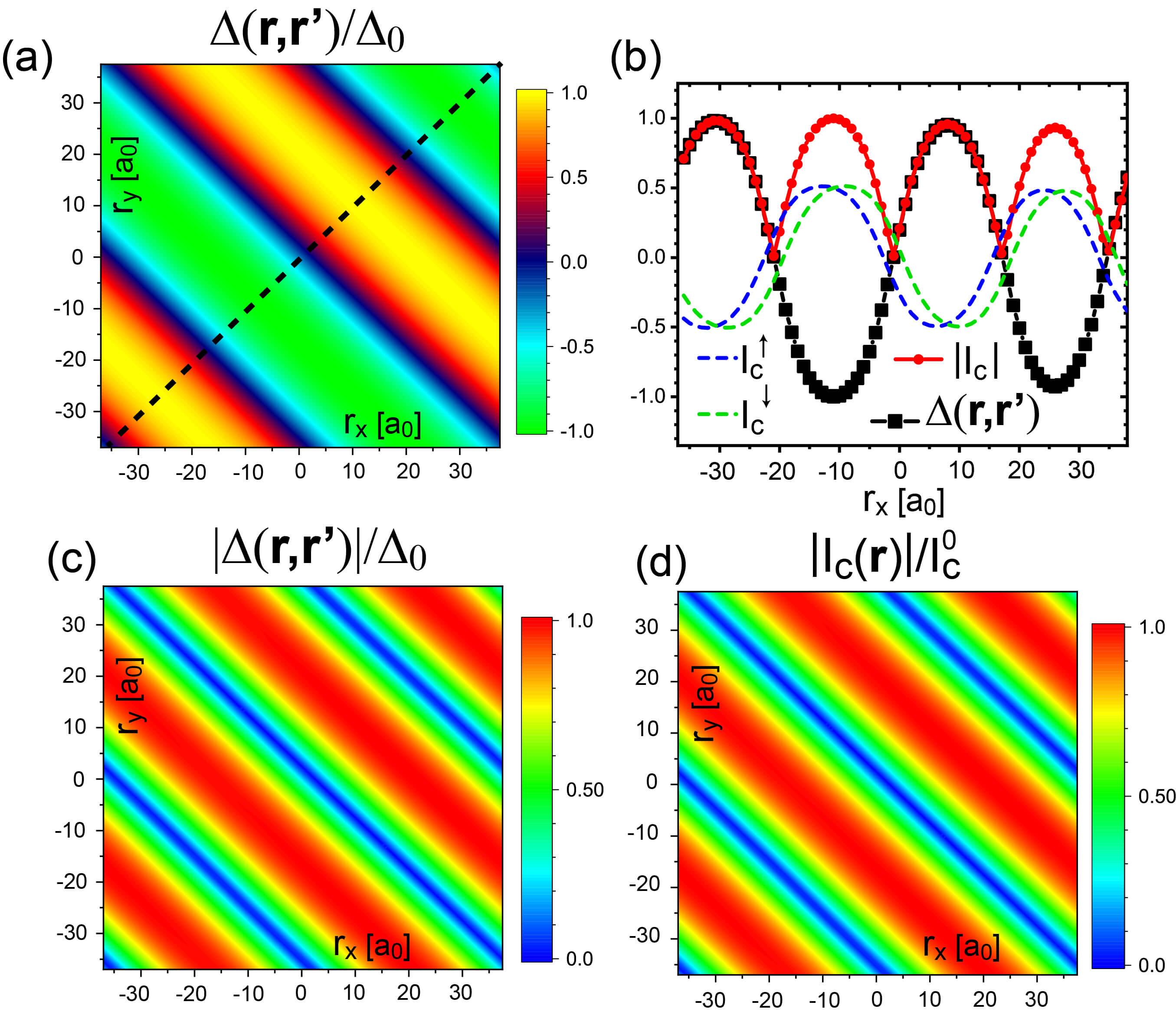}%
 \caption{(a) $\Delta({\bf r,r^\prime})$ in the FFLO phase for $g\mu_BH/2=0.1t$. (b) Linecut of $\Delta({\bf r,r^\prime})$ and $I_c=I_c^\uparrow+I_c^\downarrow$ along the black dashed line in (a) with $\Delta_{tip}=\Delta_0$. (c) Modulus of $\Delta({\bf r,r^\prime})$ and (d) $I_c$ for a $(2 \times 1)/(1 \times 2)$ tip.}
 \label{fig:FFLO}
 \end{figure}

The magnetic-field induced Fulde-Ferrell-Larkin-Ovchinikoff phase represents another example for a phase in which the SCOP exhibits characteristic spatial oscillations. Such a phase might be realised in the heavy fermion superconductor CeCoIn$_5$ \cite{Bia03,Mat07,Radovan03}, which was argued to possess a superconducting $d_{x^2-y^2}$-wave symmetry \cite{All13,Dyke14}. Solving Eq.~(\ref{eq:OP}) for $\Delta_{{\bf r}{\bf r}^\prime}$ in the presence of a magnetic field, we find that the SCOP shows sinusoidal spatial oscillations accompanied by a sign change [see Figs.~\ref{fig:FFLO}(a) and (b)], reflecting a non-zero center-of-mass momentum of the Cooper pairs in the FFLO phase \cite{Yan09,Liu12}.
As current JSTS experiments can measure only the magnitude of $I_c$, but not $I_J$ itself, we compare in Figs.~\ref{fig:FFLO}(c) and (d) the spatial structures of the modulus of $\Delta({\bf r,r^\prime})$ and $I_c$, which show very good agreement. Moreover, a linecut of the SCOP and $I_c$ in Fig.~\ref{fig:FFLO}(b) reveals that the sign change in $\Delta({\bf r,r^\prime})$ leads to a characteristic $|\sin(kr)|$ structure in $I_c$. Thus the measured $I_c$ does not only reflect $|\Delta({\bf r,r^\prime})|$, but its spatial form can also detect sign changes in the SCOP. These results, taken together, show for the first time that it is possible to detect the presence of the FFLO phase, and reveal its much anticipated spatial SCOP structure via JSTS. Finally, we note that the Zeeman-splitting of the spin-$\uparrow$ and spin-$\downarrow$ bands  \cite{Vor05} also possesses a counterpart in $I_c$: its spin-$\uparrow$ ($I_c^\uparrow$) and spin-$\downarrow$ ($I_c^\downarrow$) contributions, shown by the dashed blue and green lines in Fig.~\ref{fig:FFLO}(b), respectively, are spatially split due to the SCOP's finite center-of-mass momentum.

\begin{figure}
\includegraphics[width=8cm]{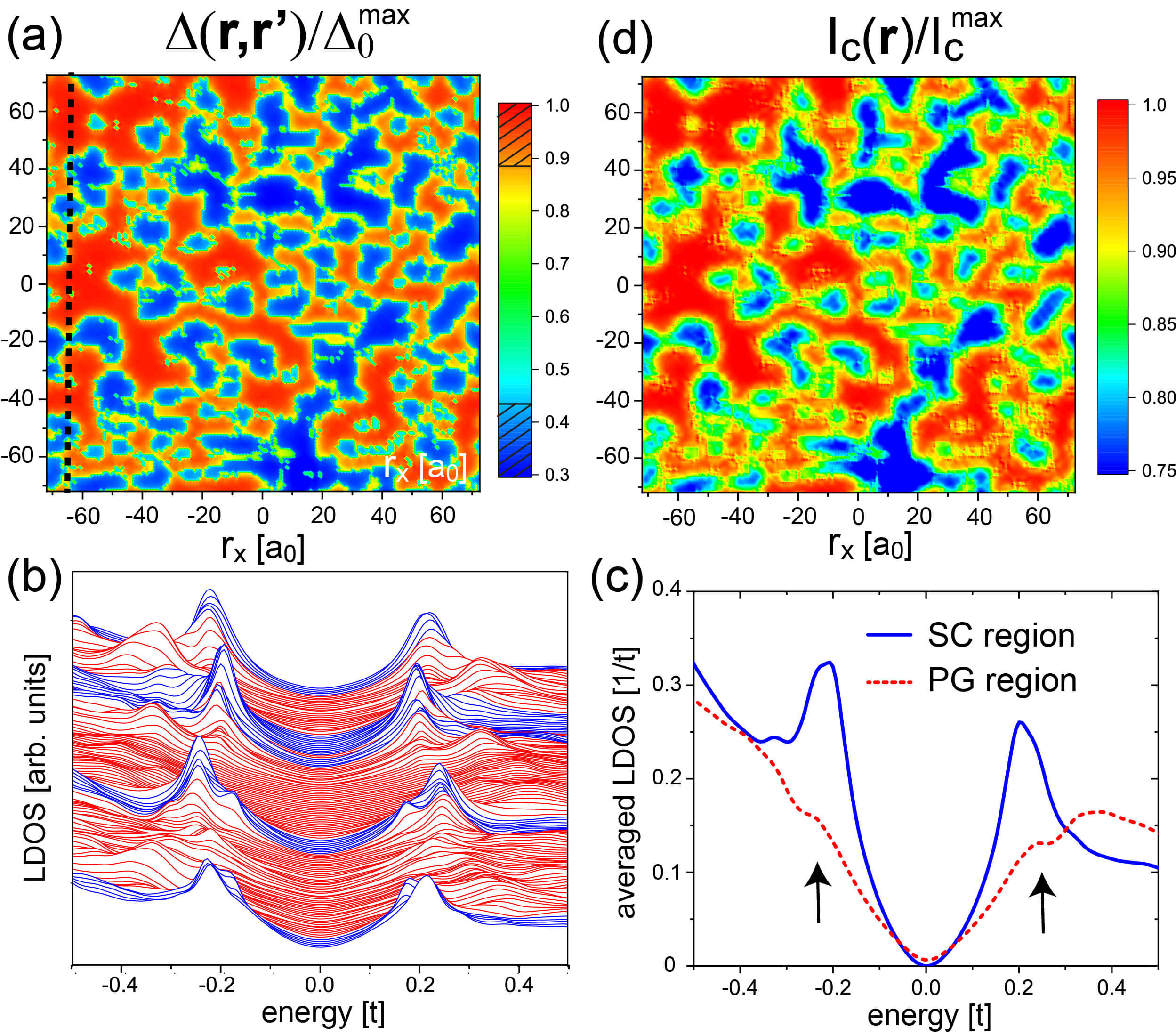}%
 \caption{Theoretical (to-scale) model of the gap-map shown in Fig.1{\bf a} of Ref.~\cite{Lang02}). (a) Spatial plot of $\Delta({\bf r,r'})$ with $\tau_{ph}^{-1}=0.05t/\hbar$ for the PG regions (red) and $\tau_{ph}^{-1} \rightarrow 0$ for the SC regions (blue). (b) LDOS along the black dashed line in (a). Blue (red) curves correspond to the SC (PG) regions. (c) Averaged LDOS (see text) in the PG and SC regions. (d) Spatial plot of $I_c({\bf r})$ for a $(2 \times 1)$/$(1 \times 2)$ tip.}
 \label{fig:pseudogap}
 \end{figure}

As the measurement of the critical current is a local probe, we expect that a non-zero $I_c$ is measured as long as a system exhibits local superconducting correlations. As such, JSTS possesses the potential to probe the nature of the pseudo-gap, and in particular, its proposed origin arising from phase-incoherent superconducting correlations (precursor pairing) \cite{Eme95,Fra98,Chen05,Wul09}. To explore this possibility, we start from the observations of conventional STS experiments \cite{Lang02,How01,Sch11} which reported the existence of a heterogeneous, domain-like structure in the underdoped cuprates at $T \ll T_c$: in one type of domain, the LDOS exhibits all the traits of a $d_{x^2-y^2}$-wave superconductor with well defined coherence peaks (the SC region), and one in which the gap appears larger than in the superconducting regions, but in which the coherence peaks are significantly broadened (the PG region). Below, we model the pseudo-gap as arising from phase incoherent superconducting correlations \cite{Chen05,Wul09}, as characterized by a finite phase-coherence time $\tau_{ph}$, such that $\Delta({\bf r,r^\prime})$ in the PG regions should be interpreted as a measure for the bond superconducting correlations, rather than a phase-coherent bond order parameter \cite{Chen05,Wul09}. To investigate the form of $I_c$ in such a heterogeneous system, we created a theoretical real-space (to-scale) model of the experimental gap-map shown in Fig.1{\bf a} of Ref.~\cite{Lang02}, which reflects the existence of the two domains. The self-consistently computed OP shown in Fig.~\ref{fig:pseudogap}(a) reproduces well the spatial structure of the experimental gap-map (for details, see SM Sec.III) . Note that in a heterogeneous system, the theoretical OP is in general not identical to the gap which is experimentally determined from the position of the coherence peaks.
In Fig.~\ref{fig:pseudogap}(b), we show a line cut of the LDOS [along the black dashed line in Fig.~\ref{fig:pseudogap}(a)], which exhibits the characteristic evolution of the LDOS between the SC and PG regions also found in STS experiments (see Fig.3{\bf a} of Ref.\cite{Lang02}). To directly compare our results with the spatially averaged experimental $dI/dV$ data (see Fig.3{\bf b} in Ref.\cite{Sch11}), we present in Fig.~\ref{fig:pseudogap}(c), the LDOS spatially averaged over the SC and PG regions along the line cut in Fig.~\ref{fig:pseudogap}(a) [these regions are defined by an OP that lies within the shaded blue (SC) or red (PG) regions of the legend in Fig.~\ref{fig:pseudogap}(a), respectively]. Our results reproduce a series of characteristic traits exhibited by the experimentally averaged data (see Fig.3{\bf b} in Ref.~\cite{Sch11}): (i) while the gap in the PG region is larger than in the SC region, the coherence peaks in the PG region are smeared out,(ii) the LDOS in the PG regions (red dashed line) exhibits shoulder-like features (see arrows) at the energies of the coherence peaks in the SC regions, (iii) the shoulder-like feature is more pronounced at positive than at negative energies, and (iv) the spectra exhibit overall a strong particle-hole asymmetry. All of these results are in good agreement with the experimental findings, thus supporting the validity of the model employed here. In Fig.~\ref{fig:pseudogap}(d), we present a spatial plot of $I_c$ obtained for the heterogeneous system shown in Fig.~\ref{fig:pseudogap}(a). We again find that $I_c$ spatially images the OP, and that despite the incoherent nature of the PG region, the Josephson current in these regions is non-zero (though $I_c$ in the PG regions decreases with decreasing $\tau_{ph}$). This implies that $I_c$ indeed reflects the presence of local superconducting correlations, rather than global phase coherence. In contrast, if the PG were to arise from non-superconducting correlations or order parameters, such as charge- or spin-density wave correlations, $I_c$ would vanish. Thus, a non-zero measurement of $I_c$ in the PG region would provide strong evidence for precursor pairing as its origin.

In conclusion, we have shown that JSTS in unconventional $d_{x^2-y^2}$-wave superconductors provides unprecedented insight into the spatial form of the SCOP not only near defects but also in the magnetic-field induced FFLO phase. While it is possible to image the bond SCOP for sufficiently small JSTS tip sizes, we find more generally that the spatial form of $I_c$ images that of the SCOP averaged over the size of the tip.  Moreover, we showed that in the FFLO phase, JSTS does not only map the spatial variations in the magnitude of the SCOP, but also visualizes its sign change. Finally, we demonstrated that JSTS can detect the presence of phase-incoherent superconducting correlations, thus providing insight into the potential origin of the PG from precursor pairing and discriminating it from other proposed mechanisms.

\begin{acknowledgments}
We would like to thank C. Ast, J.C. Davis, and M. Hamidian for helpful discussions.  This work was supported by the U. S. Department of Energy, Office of Science, Basic Energy Sciences, under Award No. DE-FG02-05ER46225.
\end{acknowledgments}

\end{document}


\fontsize{11}{13}

\begin{center}
{\large {\bf  Supplemental Online Material for}} \\[0.5cm]
\end{center}

\title{Josephson Scanning Tunneling Spectroscopy in $d_{x^2-y^2}$-wave superconductors: a probe for the nature of the pseudo-gap in the cuprate superconductors}

\author{Martin Graham and  Dirk K. Morr}

\affiliation{Department of Physics, University of Illinois at Chicago, Chicago,
IL 60607}

\date{\today}

\maketitle

\section{Definition of Green's functions in real space}
To compute the spatial dependence of the non-local $d_{x^2-y^2}$-wave order parameter, $\Delta_{{\bf r}{\bf r'}}$, as well as the critical Josephson current, $I_c({\bf r})$, in the presence of defects, we rewrite the Hamiltonian in Eq.(1) of the main text in matrix form introducing the spinor
\begin{align}
\Psi^\dagger = \left(c^\dagger_{1,\uparrow}, c_{1,\downarrow}, \ldots, c^\dagger_{i,\uparrow}, c_{i,\downarrow}, \ldots, c^\dagger_{N,\uparrow}, c_{N,\downarrow} \right)
\label{eq:psi}
\end{align}
where $N$ is the number of sites in the $d_{x^2-y^2}$-wave superconductor, and $i=1,...,N$ is the index for a site ${\bf r}$ in the system. The Hamiltonian in Eq.(1) of the main text can then be written as
\begin{align}
H_s = \Psi^\dagger {\hat H}_s \Psi \ .
\end{align}
We next define a retarded Green's function matrix of the system via
\begin{align}
{\hat G}_{sc} (\omega+i \delta) = \left[ (\omega+i \delta) {\hat 1} - {\hat H}_s \right]^{-1}
\end{align}
where ${\hat 1}$ is the $(N \times N)$ identity matrix and $\delta = 0^+$. The non-local anomalous Green's function $F_{sc}({\bf r'}, \downarrow; {\bf r},\uparrow, \omega)$ between sites ${\bf r'}$ (with index $j$) and ${\bf r}$ (with index $i$) that enters the calculation of $I_J^\uparrow$ [see Eq.(5) of the main text] is then the given by the $(2j,2i-1)$ element of ${\hat G}_s$. The anomalous Green's function $F_{sc}({\bf r'}, \uparrow; {\bf r},\downarrow, \omega)$  involved in the calculations of $I_J^\downarrow$ is obtained from the relation
\begin{align}
F_{sc}({\bf r'}, \uparrow; {\bf r},\downarrow, \omega)=-F^*_{sc}({\bf r'}, \downarrow; {\bf r},\uparrow, -\omega)
\end{align}

Moreover, we take the anomalous Green's function of the tip, $F_t({\bf r},\uparrow; {\bf r'}, \downarrow, \omega)$, to be that of a bulk system which can be computed from the momentum space form of the anomalous Green's function,
\begin{align}
F_t({\bf k},\omega) = - \frac{\Delta_{\bf k}}{E_{\bf k}} \left[ \frac{1}{\omega-E_{\bf k} + i \delta} - \frac{1}{\omega+E_{\bf k} + i \delta}\right]
\end{align}
via
\begin{align}
F_t({\bf r},\uparrow; {\bf r'}, \downarrow,\omega) = \int \frac{d^2k}{(2 \pi)^2} \, F_t({\bf k},\omega) e^{i {\bf k}( {\bf r}- {\bf r'})}
\label{eq:AGF_tip}
\end{align}
where
\begin{align}
E_{\bf k} & = \sqrt{\varepsilon^2_{\bf k} + \Delta_{\bf k}^2} \nonumber \\
\varepsilon_{\bf k} &= -2t \left[ \cos{k_x}  + \cos{k_y} \right] - 4 t^\prime \cos{k_x} \cos{k_y} - \mu \nonumber \\
\Delta_{\bf k} & = \frac{\Delta_0}{2}  \left[ \cos{k_x}  - \cos{k_y} \right]
\end{align}

\section{Effects of Disorder in the JSTS tip}

\begin{figure}[h]
\includegraphics[width=12cm]{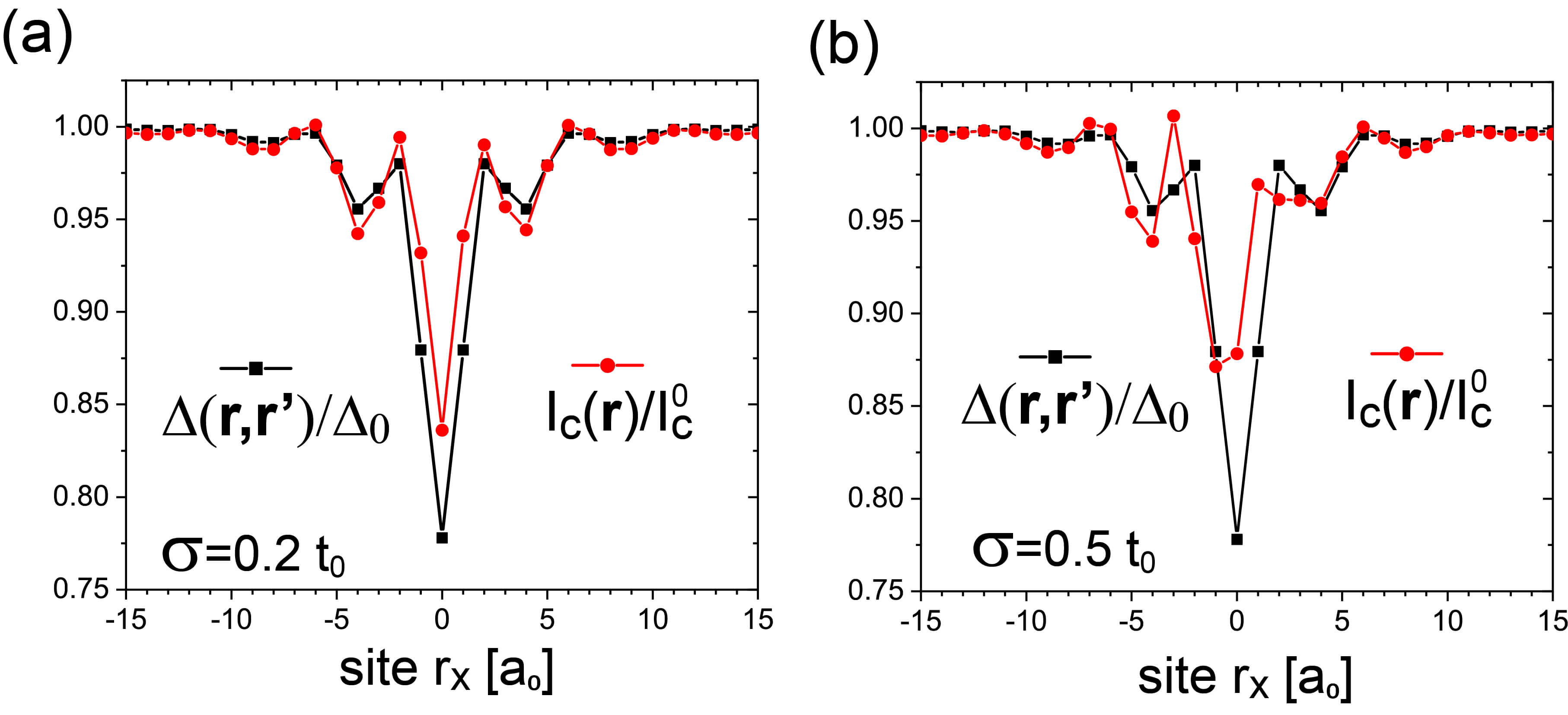}%
 \caption{Normalized $\Delta({\bf r,r'})$ and $I_c$ for a $(3 \times 3)$ tip in the presence of disorder in the tunneling amplitude with standard deviation (a) $\sigma=0.2 t_0$, and (b) $\sigma=0.5 t_0$, where $t_0$ is the tunneling amplitude in the clean case.}
 \label{fig:disorder}
 \end{figure}

In the main text, we considered highly ordered tips with identical tunneling amplitudes between the sites in the tip and the sites in the system. However, it is very likely that in the experimental case, the tip is disordered to some extent, resulting in disorder in the tunneling amplitudes. To investigate the effects of such a disorder on the spatial structure of the measured Josephson current, we consider a spatially varying tunneling amplitude, $t_0({\bf r})$. In this case, the Josephson current is given by
\begin{align}
I^\sigma_{J} &= 4\frac{e}{\hbar} \sin{\left(\Delta \Phi \right)} \sum_{ {\bf r}, {\bf r}^\prime} t_0({\bf r})t_0({\bf r}^\prime)  \int \frac{d\omega}{2\pi}n_F(\omega)
\text{Im}[F_t({\bf r},\sigma;{\bf r'},{\bar \sigma}, \omega) F_{sc}({\bf r'}, {\bar \sigma}; {\bf r},\sigma, \omega)]
\label{eq:Ieq}
\end{align}
In Fig.~\ref{fig:disorder}, we compare the normalized $\Delta({\bf r,r'})$ and $I_c$ for a $(3 \times 3)$ tip in which the tunneling amplitude possesses a standard deviation of $\sigma=0.2 t_0$ [Fig.~\ref{fig:disorder}(a)] and $\sigma=0.5 t_0$ [Fig.~\ref{fig:disorder}(b)], where $t_0$ is the tunneling amplitude in the clean case.
We find that even for the case of $\sigma=0.5 t_0$ [Fig.~\ref{fig:disorder}(b)], which represents already a significant amount of disorder, the agreement between the spatial structure of $\Delta({\bf r,r'})$ and $I_c$ is still quite good. The reason for the good agreement even in the stronger disorder case is that for the spatially extended tip considered here, the measured $I_c$ is proportional to the spatially averaged superconducting order parameter, due to the summations over ${\bf r}$ and ${\bf r^\prime}$. These summations also lead to an averaging out of the disorder effects, implying that site-dependent disorder has less of an impact for spatially extended tips. This demonstrates that the ability of $I_c$ to spatially image $\Delta({\bf r,r'})$ persists even in the presence of significant tip disorder.

\section{The Pseudo-Gap region: a phase incoherent $d_{x^2-y^2}$-wave superconductor}

It has previously been argued that the pseudo-gap regions of the cuprate superconductors can be described as a phase-incoherent $d_{x^2-y^2}$-wave superconductor, where the finite phase coherence time, $\tau_{ph}$ is introduced phenomenologically into the Green's function matrix via (with $\gamma = \hbar / \tau_{ph}$)
\begin{align}
{\hat G}_{pg} (\omega+i \gamma) = \left[ (\omega+i \gamma) {\hat 1} - {\hat H}_s \right]^{-1}
\end{align}
As before, the gap in the pseudo-gap region is computed self-consistently using
\begin{align}
\Delta^{pg}_{{\bf r}{\bf r}^\prime} & =-\frac{V_{{\bf r}{\bf r}^\prime}}{\pi}\int_{-\infty}^\infty d\omega \, n_F(\omega) \text{Im}[F_{pg}({\bf r},{\bf r'},\omega)]
\end{align}
A finite $\tau_{ph}$ immediately leads to a broadening of the coherence peaks, as shown in Figs.~3(b) and 3(c)of the main text, where we used $\gamma=0.05t$ in the PG regions, and $\gamma =0.005t $ in the SC regions. In order to achieve that the coherence peaks in the PG regions are located at higher energies (while at the same time being broadened), we increased the effective pairing interaction in the PG regions over that in the superconducting regions. As a result, the OP in the PG regions is larger than in the superconducting regions [see Figs.~3(b) and 3(c) of the main text].